\documentclass[twocolumn,10pt]{asme2ej}
\usepackage{amsmath}
\usepackage{epsfig} 

\title{A short review on techniques for processes and process simulation of scaffold-free tissue engineering}

\author{Ali Bakhshinejad
    \affiliation{
	Graduate Research Assistant\\
	Department of Mechanical Engineering\\
	University of Wisconsin-Milwaukee\\
	Milwaukee, Wisconsin 53211\\
    Email: bakhshi3@uwm.edu
    }	
}

\begin{document}

\maketitle    

\begin{abstract}
{\it The invention of three-dimensional printers has led to major innovations in tissue engineering. They have enabled the printing of complex geometries such as those that occur in natural tissues, that were not possible with traditional manufacturing techniques. Tissue engineering in particular deals with printing bio-compatible material that may be infused with live cells. Thus additional complexity is incurred because the live cells can migrate and proliferate and thus change the printed geometry. One of the important issues is the prediction of geometry and possibly mechanical properties of the steady state tissue. In this short review, we will provide an overview of different tissue engineering processes that are currently available. Furthermore, we will review two important techniques, namely, Cellular Potts Model (CPM), and Cellular Particle Dynamics (CPD) that have been used to predict the steady state of printed tissue.
}
\end{abstract}



\section{Introduction}

Tissue Engineering is a young and fast growing research field whose objective is to create in vitro 3-D cell constructs, which, once placed in vivo environment, will differentiate to resemble natural systems \cite{Williams2005,Saltzman2004}. This field is a response to a rising life expectancy combined with an increasing number of diseases which has cause an increasing demand for organ transplants. As an example: from July 2008 to July 2009, about 105,000 people in the United States were waiting for an organ transplant, with less than a third receiving it \cite{Organdonor2011}. Tissue engineering has the potential to address this problem. In addition, it can be used to fabricate tissue samples for {\it in vitro} disease models which can be used to accelerate drug development/delivery research \cite{Huh2010a,Gunther2010}.\\

\begin{figure*}
\centering\includegraphics[width=0.8\linewidth]{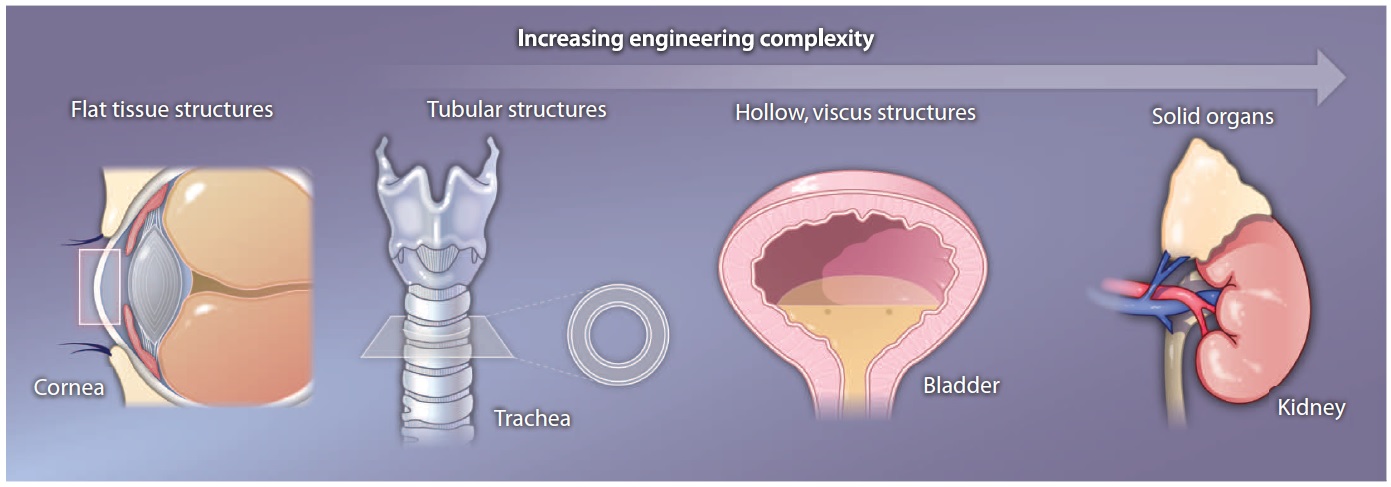}
\caption{Four levels of structural complexity for organs. As the structure and metabolic function of the tissue or organ increases the complexity of a tissue engineering approach increases as well. This figure is adopted from Atala et al. \cite{Atala2012} }
\label{fig:OrganComplex}
\end{figure*}

To reconstruct any organ, one has to engineer to the target application. Therefore, categorization of organs based on structural complexity is important. Generally, organs can be categorized into four levels of structural complexity: flat tissue structures tubular structures, hollow and viscous structures, and solid organs (Figure \ref{fig:OrganComplex}  ).\\

\begin{figure}
\centering\includegraphics[width=0.8\linewidth]{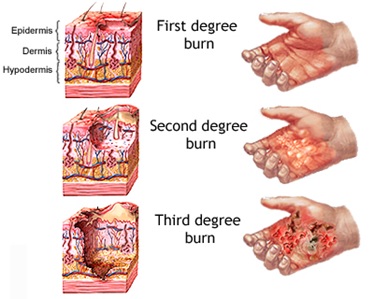}
\caption{Burn Degrees. The depth of the burn determines its severity. \cite{Med}  }
\label{fig:BurnLevel}
\end{figure}

Flat tissues, which have the simplest architectural structure, can be describe as sheets of cells, constructed from multiple layers of one cell type. These structures do not have any inter-layer and intra-layer vascular network. Tissues in this category are thin enough where cells can feed and eject their waste just using diffusion. The cornea and skin belong to his category.\\
The effects of substantial loss of skin surface area due to bruns are quite detrimental due to increased risk of infection and fluid loss. This condition is a priority in tissue engineering. Based on the skin burn level (Figure \ref{fig:BurnLevel}), several methods are in use to engineer adequate skin. For first-degree skin burns, normal skin cells are  harvested from the patient and are then applied to the burn area \cite{Gerlach2011,Gerlach2011a}. Yet another treatment method involves harvesting the patient’s skin cells and grafting them onto a scaffold and subsequent implantation over the burn area \cite{Carsin2000,Chau2013}. 
Engineering the skin for higher degrees of burn or so called full-thickness burns, where the vascular system is damaged, is still a challenge for researchers \cite{Atala2012}.\\

	For engineering tubular structures, the traditional method is to locate cells on scaffold sheets and form them into circular shapes. These structures' functionality is limited to carrying air or fluid through the body and generally consist of two different cell types. Since they do not have any vascular network and are thin enough to allow diffusion, their construction is not that challenging.\\
    
	Hollow, viscous structures placed into third level of complexity. Organs that fall into this category (i.e. bladder) are constructed from two type of cells same as tubular structures but unlike tubular structures these group have higher functionality as well as metabolism, i.e., the can have vascular structures for nutrient and waste transport.\\

	The last category, solid organs, such as the kidney, liver and heart are the most complex structures. Traditional treatment for these type of organs, depends on type of disease, consist of either temporary management with drugs/devices or whole-organ transplantation. In case of transplantation even those small portion of patients that receives the needed organ, they have to take immunosuppressant drugs to prevent the graft from being rejected by the immune system, more likely for lifetime periods \cite{Koch2013}.\\

Tissue engineering can be approached in one of three different ways: bio-mimicry, autonomous self-assembly and mini-tissue building blocks \cite{Murphy2014}. Bio-mimicry methods or the so called scaffold-base methods use bio-materials to construct a scaffold. The scaffolds provide temporary extra-cellular matrix (ECM) for tissue during the maturation time \cite{Almeida2014}. Scaffolds must provide a bio-compatible, biodegradable micro-environment with highly porous structures to maintain regeneration of the new tissue out of the cells. For this approach to succeed, an understanding of the biological micro-environment is needed \cite{Murphy2014}. The process of maturation changes the structure over time. Therefore, software tools have been developed to predict the steady state of the tissues.  Computer Aided system for Tissue Scaffolds (CASTS)  \cite{Naing2008} and Computer Aided Design of Scaffolds (CADS) \cite{Rainer2012} are two examples of this method. Other groups are researching other problems addressing bio-materials, cell biology, biophysics and medicine.\\

 Self-assembly (scaffold-free) is the autonomous organization of components, from an initial state into a final pattern or structure without external intervention \cite{Jakab2010,Tasoglu2013}. Self-assembly methods address the problem of biodegradability of scaffold-based methods \cite{Kasza2007}. Using cells as building blocks is the primary choice for self-assembly methods\cite{Kasza2007}. On the other hand, in order to scale the process to level that makes it commercially viable, using tissue spheroids as building blocks has been suggested\cite{Rezende2013}. 

The last approach, mini-tissue building blocks, is relevant to both previous strategies. Tissues and organs consist of smaller building blocks or mini-tissues\cite{Murphy2014}. Mini-tissues can be designed and fabricated using either of the previous methods. In this method, self-assembling cell spheres are assembled into a micro-tissues \cite{Rezende2013}. In the next step, high-resolution reproduction of a tissue unit are placed and allowed them to self-assemble into a functional macro-tissue \cite{Murphy2014,Blakely2014}.

The technology of 3-D printers, first presented by Charles W. Hull \cite{Hull1986}, inspired tissue engineering like other engineering fields. The main idea of 3-D printers, printing thin layers of material, one at the time, to form a solid 3-D structure was used to develop bio-printers. Current generation bio-printers belong to two main classes: Laser assisted printers\cite{Koch2013,Odde2000a,Guillotin2010a,Gruene2011} and laser-free printers\cite{Ringeisen2006,Faulkner-Jones2013,BINDER2011}. 

Like any other production process, bio-printing needs to have a process plan that has to be executed to achieve the desired tissue structure and functionality. However, unlike traditional 3-D printing,  the manufactured tissue will mature through processes such as cell division, cell migration, and cell sorting. This changes both the shape and characteristics of the tissue. Therefore, the process plans, which may involve bio-material deposition paths and cell placement plans have to account for changes due to the maturation processes before hand. 

In this paper we focus on scaffold-free tissues engineering processes. We begin with a discussion of bio-printing process planning. We discuss the design process as well as techniques for simulation-based design verification. We then proceed to discuss techniques for scaffold-free tissue engineering. Next, we describe various maturation techniques. Finally, we summarize the state of the art and directions for future research.

\section{Bio-printing Process Plan}
 The process-planing steps for bio-printing involves a) Tissue imaging, b)Computer-aided design (CAD), c)cell selection d) simulation-based prediction and correction of the design.  

\begin{figure*}
\centering\includegraphics[width=0.8\linewidth]{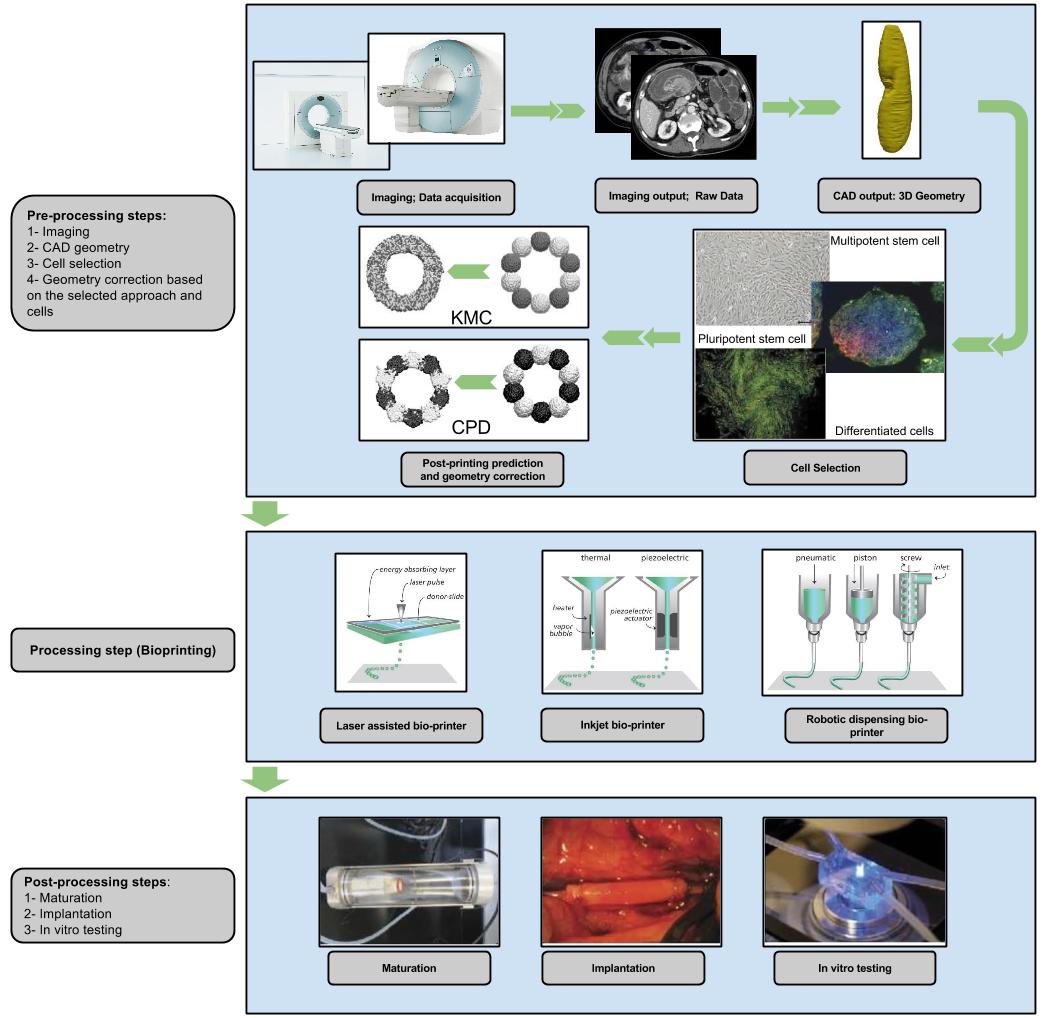}
\caption{ A typical process for bioprinting of tissue/organ. Imaging of the damaged tissue and environment using different modalities such as MRI or CT is the first step of the process which generates raw images. Raw image data is used to reconstruct a three-dimensional geometry of the organ using either commercial (e.g. Mimics \cite{Materialise}) or open source (e.g. CGAL \cite{cgal:h-af-14b}) softwares.  Selecting cell type depends on the project requirement will provides mechanical properties that will be used for prediction and geometry correction purposes. After having the initial data (e.g. 3D geometry and mechanical properties of cells) using one of the available simulation algorithm the final shape of the organ can be predicated based on the initial locations of the cells. Applying required corrections the corrected initial locations can be calculated and using any of different bio-printers, one can print out the organ. Tissues that has been printed using this method required maturation before be eligible for implantation or even in vitro tests. Cell selection images and post-processing images are reprinted from Murphy \cite{Murphy2014}. KMC and CPD images are reprinted from Flenner et al. \cite{Flenner2012}. Bio-printer images are reprinted from Mala et al. \cite{Malda2013}.  }
\label{fig:Bio-printing_steps}
\end{figure*}

\subsection{Imaging}
An essential requirement for reproducing the complex and heterogeneous structure of any highly porous media, including functional tissues and organs, is a comprehensive understanding of composition and organization of their components.  Therefore, imaging is the first step in tissue engineering pipeline. The computed tomography (CT) and magnetic resonance imaging (MRI) are two of the most popular noninvasive imaging modalities that have been used by tissue engineers for a better understanding of structures at different levels. Each of these methods generate voxelized volumetric data.

The CT imaging method uses restricted x-ray beams called fan beams in order to produce cross-sectional images. Several different generation of CT imaging techniques have been developed since the early invention of Allan M. Cormack and Godfrey N. Hounsfield \cite{Nobel1979}. Improvements over several generations have resulted in increased scan speed and spatial resolution. In CT imaging, the x-ray source rotates around the object and generates fan beams at different angles. The signal detectors measure the transmitted beam intensity at every given angle. The attenuation coefficient map over the cross-section are then obtained from the transmitted beam intensity data using inverse X-Ray transform.
 \cite{Prince2006}.  
    
	The second method, the MRI, generates high spatial resolution raw data using the principles of nuclear magnetic resonance. In this method, the whole object is magnetized with a main magnetic field (typically 1.5 T) which causes alignment of Hydrogen atoms in the field. After alignment, a secondary magnetic field applies on the object, utilizing three additional gradient magnetic field voxel of interest for imaging can be selected. Using radio frequency (RF) coils, selected voxel’s nuclei will be excited, and the free induction decay of atoms can be sensed after turning off the excitation radio frequencies. The MR images are produced by analyzing the measured raw data \cite{Prince2006}.

The next step is the extraction of three-dimensional geometry and its visualization from the generated raw image data, is one of the most important steps in the pre-processing analysis \cite{Park2011}.  There are three sub-steps in the extraction process: pre-processing, segmentation, and visualization. Pre-processing step involves image alignment (parametric/deformable) for compensation of the movements between consecutive slices and extracting the correspondence between image features \cite{baghaie2014fast}, slice interpolation for asymmetry correction of different resolutions along the three axes  \cite{Baghaie2014}, noise reduction \cite{baghaie2014sparse} etc. For segmentation step, due to different variables like, anatomical objects of interest, large variation of their properties in images, different medical imaging modalities there is no general and unique solution. Because of that, different mathematical methods and algorithms have been developed in the last two decades to address these problems \cite{Park2011,Elnakib2011,Sharma2010, baghaie2014state}. Bartz and Preim \cite{Bartz2011} divided the visualization step into two categories: direct and indirect methods. In indirect volume rendering, or iso-surface rendering, a polygonal representation of the volume dataset is used for visualization, while in direct volume rendering,  the voxel data and not the intermediate representation is used for visualization \cite{Bartz2011,Preim2007h,Hadwiger2006}. Since patient specific models are not reliable for tissue engineering, due to disease or injury, a detailed anatomical model of organ can be used for next steps. Having an accurate three-dimensional model of the imaged organ, one can use a slicing method (dividing the 3D geometry to thin horizontal slices) in order to directly use in bio-printers or prediction models (CPM, CPD and etc.).

\subsection{Prediction and correction of the design}

	In the traditional 3-D printing process, the shape obtained does not mutate over time. Therefore, the process-planning step mainly involves slicing of geometry, and 2-D tool-path generation for both part and support structures (depending on the 3-D printing process type). Bio-printing process on the other hand involves live cells. These cells respond to environmental stimuli, migrate, proliferate and therefore can completely change the structure of the printed shape over a period of time.

    Therefore, process planning in such an environment involves the inverse problem of computing printing plans based on the desired shape of the final structure given the models of various post-printing reactions and cell rearrangements. Two most known cell rearrangements (Post-processing steps) are cell sorting and tissue fusion \cite{Perez-Pomares2006}.

\begin{figure}
\centering\includegraphics[width=0.8\linewidth]{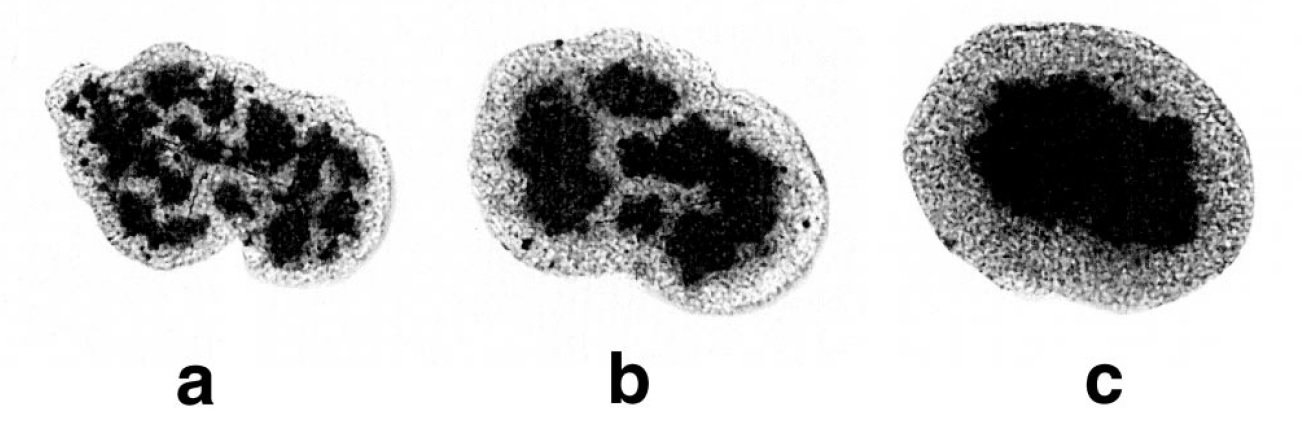}
\caption{ Sorting out of chicken embryonic pigmented epithelial cells (dark) from chicken embryonic neural retinal cells (light). The average aggregate size is 200 $\mu m$. At the end of sorting, neural retinal cells preferentially wet the external tissue culture medium surrounding the aggregates. a, b, and c correspond to 17 h, 42 h, and 73 h after initiation of sorting, respectively. \cite{Beysens2000} }
\label{fig:cellsorting}
\end{figure}

Cell sorting process is a common self-assembly mechanism providing cellular compartments discretization during various developmental processes \cite{Jakab2010,Perez-Pomares2006,Armstrong1989}. Cell sorting happens when the aggregates contain two or more different cell types. In such cases, sorting starts from an initially disordered array of cohering cells into a situation in which cells are organized into homogeneous tissue domains \cite{Armstrong1989, Graner1992}. Cell sorting has been shown experimentally in 2-D \cite{GARROD1973} and 3-D \cite{Steinberg1970} (Figure \ref{fig:cellsorting}) \cite{Graner1992}.

Experiments have found that differences in intercellular adhesivity determines the final state of cell sorting with the cell configuration rearranging continuously in order to reach to the global minimum overall surface energy \cite{Steinberg1963,PHILLIPS1978}. Surface energy between different cells is the main driving force for cell sorting \cite{Graner1992}. These energies include differences between surface energy for boundaries between monotypic and heterotypic cells, and with the external medium (edge). Different studies have been reported that the cell sorting process is governed by the same physical forces driving phase separation of immiscible liquids \cite{Beysens2000,Foty1994,Duguay2003}.

	Tissue Fusion is another self-assembly mechanism in which different cell aggregates merge into each other and shape a bigger homogeneous tissue. This process involves  mutual adhesiveness between the different cell types involved in the process. The cells involved must be able to cross-adhere but not repel each other or tend to remain separated because of their low mutual adhesivity. The fusion effect has been reported and studied in detail by different groups both in vitro and vivo environment \cite{Perez-Pomares2006}.
    
	Different mathematical models have been proposed to predict post-printing reactions and rearrangements. The Gastrulation was modeled by Odell and coworkers \cite{Odell1981} by modeling cells as a collection of coupled viscoelastic elements. Palsson and Othmer studied the effect of individual cell movement on motion of an aggregate of cells by modeling cells as deformable viscoelastic ellipsoids \cite{Palsson2000}.

Cellular Potts Model (CPM) represented cells as a collection of contiguous spins, defined on a discrete lattice \cite{Glazier1993}. Kosztin and coworkers represented cells as a continuum material object with defined volume and adaptive shape (the model called Cellular Particle Dynamic (CPD))  \cite{Flenner2008}, these are some of the many mathematical models that has been developed during the last three decades. CPD\cite{Flenner2008} and CPM \cite{Graner1992,Glazier1993} are the most well-known and accurate models, which we briefly described each model in following sections.

\subsubsection{Cellular Potts Model (CPM)}

Representing a single region in space by multiple regular lattices is a way to overcome the complexity of the subcellular level complexity that was introduced by Glazier-Graner-Hogeweg (GGH) model \cite{Glazier2007}. This model was developed based on cell sorting experiment on top of the available large-q potts model in order to simulate the cell sorting in two and three dimensions \cite{Sulsky1984,Steinberg1975,Anderson1984,Srolovitz1984,Glazier1990}. 
	The GGH typical application domain is to model soft tissues with motile cells at a single cell resolution. Due to the general formulation of the model in GGH, a general cell may represent a biological cell, a subcellular compartment, a cluster of cells or a piece of non-cellular material \cite{Swat2012}. In this model, each generalized cell is a domain of lattice pixels in the cell lattice that shares a common index as the cell index  $ \sigma$. In the case of the biological cell, the cell is defined as a cluster of generalized cells in order to address the cell compartments, shape, polarity etc. \cite{Glazier2007,Swat2012,Walther2004,Walther2005}.
The model (Equation  \ref{eq:HGGH}) represents cells as patches of identical lattice indices  $ \sigma (\overrightarrow{i})$ on a lattice, where each index uniquely identifies. Each cell has an associated cell type, $\tau$. And $J(\tau ,\tau')$ represents the surface energy between spins of type $\tau$ and $\tau'$. $\lambda$ stands for Lagrange multiplier specifying the strength of the are constraint, $A(\sigma)$ is the current area of cell $\sigma$, and $A_{target} (\sigma)$ shows its target area. And finally, $\delta (x,y)$ is the Kronecker delta \cite{Graner1992,Glazier2007,shirinifard2012vascular}.

\begin{align}
\mathcal{H}_{GGH} &= \sum_{\overrightarrow{i},\overrightarrow{j} \hspace{1mm} neighbors} J(\tau(\sigma(\overrightarrow{i})), \tau(\sigma(\overrightarrow{j})))(1-\delta(\sigma(\overrightarrow{i}),\sigma(\overrightarrow{j}))) \nonumber\\
&+\sum _\sigma \lambda(A(\sigma) - A_{target}(\sigma))^2 \label{eq:HGGH}\\
\Delta \mathcal{H}_{total}& = \Delta \mathcal{H}_{chemotaxis}+\Delta \mathcal{H}_{GGH}\nonumber
\end{align}

\begin{figure}
\centering\includegraphics[width=0.8\linewidth]{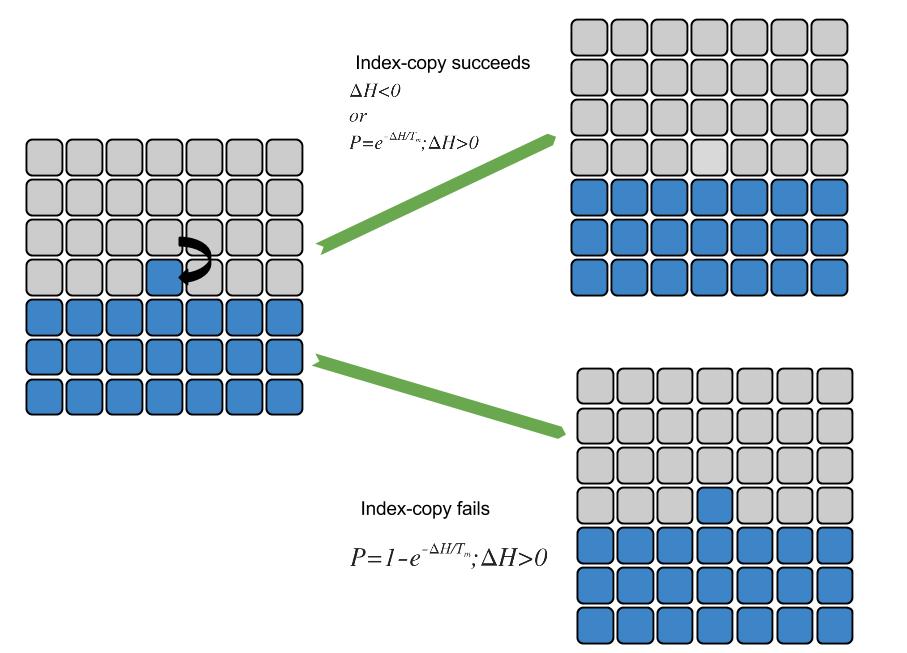}
\caption{ Probability of acceptant or rejection of any spins in CPM model can be calculated as shown \cite{Mortazavi2013}. }
\label{fig:CPM_Spin}
\end{figure}

In randomly selected target sites in the cell lattice, algorithm uses a stochastic modified Metropolis algorithm \cite{Cipra1987} to model the cells dynamic with a series of index-copy attempts. The algorithms accepts or rejects the attempts for two neighboring cells ($\sigma_{\overrightarrow{i}}$  and $\sigma_{\overrightarrow{i'}}$) with probability $P(\sigma_{\overrightarrow{i}} \rightarrow \sigma_{\overrightarrow{i'}})$ given by the Boltzmann acceptance function (Figure \ref{fig:CPM_Spin}):\\

\begin{equation}
P(\sigma_{\overrightarrow{i}} \rightarrow \sigma_{\overrightarrow{i'}}) =
\left \{
\begin{matrix} 
 1 & :\quad \Delta \mathcal{H}\le 0 \\ 
 e^{ -\frac { \Delta \mathcal{H} }{ T_{ m } }  } & :\quad \Delta \mathcal{H}>0 
\end{matrix}
\right.
\end{equation}

Where $\Delta \mathcal{H}$ is the change in the effective energy if the copy occurs and $T_m$ is describing the amplitude of cell-membrane fluctuations. For the given generalized cell, the average value of $ \Delta \mathcal{H} / T_m$  determines the amplitude of fluctuations of the cell boundary, high values for solid, barely or non-motile cells and low for a high degree of cell motility and rearrangements \cite{Swat2012}. 
	This model has been adopted and applied to simulate different biological and non-biological phenomenons like: Tumor growth \cite{Szabo2013}, Gastrulation \cite{Calmelet2010}, Angiogenesis \cite{vanOers2014} and Wetting \cite{Mortazavi2013}.  

\subsubsection{Cellular Particle Dynamics (CPD)}

	With modeling individual cells as a set of interacting cellular particles (CPs), where initially introduced by Newman \cite{Newman2005} for modeling multicellular system using Subcellular Element Model (SEM), Flenner et al. \cite{Flenner2008} were able to simulate the motion and self-assembly of large cell aggregates. This method, follows the three-dimensional spatial trajectories of individual cells by integrating numerically the corresponding equations of motion.

\begin{figure}
\centering\includegraphics[width=0.8\linewidth]{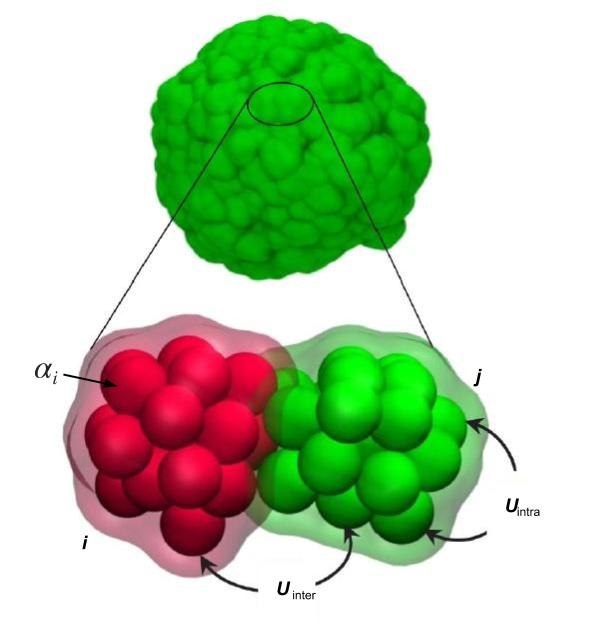}
\caption{ Close-up view of two interacting cells (bottom) in a model spheroidal multicellular aggregate (top). The CPs in contact are colored differently and labeled as i and j. $U_{intra}$ and $U_{inter}$ represents the intracellular and intercellular interaction energies between CPs \cite{Kosztin2012}.  }
\label{fig:CPD}
\end{figure}

For introduction purposes, consider a system with constant number N of cells in three spatial dimensions, with each cell being composed of M elements. With labeling each individual cell by an id, $i \in (1,N)$, and an element in cell i by $\alpha _i \in (1,M)$ (Figure \ref{fig:CPD}). The precise form of the interaction between CPs is governed by the detailed biology of the cell, whose complexity of the system makes the determination of this interaction potential practically impossible. So, assuming absent of chemical signaling, the position vector of element $\alpha _i$ is changing in time according to three processes: (1) fluctuations in the dynamics of the cellular cytoskeleton; (2) an elastic response to intracellular biomechanical forces; and (3) an elastic response to intercellular forces \cite{Newman2005}.  Therefore, considering this simplification, CPs interact can be modeled by the Lennard-Jones (LJ) potential energy \cite{Jones1924}: 
    
\begin{equation}
V_{LJ}(r;\varepsilon , \sigma) = 4 \varepsilon  [(\frac{\sigma}{r})^{12}-(\frac{\sigma}{r})^6]
\label{VLJ}
\end{equation}
        
    Where r is the distance between the CPs, $\varepsilon $ is the energy required to separate the CPs and $\sigma$ stands for the diameter of a CP. Considering two CPs with different cell id (belonging to different cells), the intercellular interaction potential can be calculate from equation \ref{VLJ} ($U^{inter} (r)= V_{LJ}  (r;\varepsilon _2,\sigma _2)$) \cite{Flenner2008}. If two CPs have same cell id (belong to the same cell), the intracellular potential can be calculate as:
    
\begin{equation}
U^{intra} (r)= V_{LJ}  (r;\varepsilon,\sigma) + \frac{k}{2} (r - \xi)^2 \theta(r-\xi)
\end{equation}

  Where $\theta(r)$ the Heaviside step function and k is stands for elastic constant \cite{Flenner2012}. Keeping in mind that instead of LJ potential energy one can use any other potential that has a repulsive core and a short range attractive part as in SEM a Morse potential is used to describe the interaction between two subcellular elements \cite{Newman2005,Flenner2012}. In addition of mentioned potential energies, the interaction between each CP and cytosol (its highly viscous environment) can be described by a friction force $F_f= - \mu \dot{r}_\alpha$ ($\mu$ is friction factor and the dot is time derivative).  Finally, with assuming that inertia can be neglected, then the corresponding Langevin equation of motion for the ith CP in the \textit{n}th cell can be written as
    
\begin{equation}
\mu \dot{r}_{i,n}(t) = - \nabla _{i,n} U+ f_{i,n}(t)
\end{equation}
    
\begin{figure}[t]
\centering\includegraphics[width=0.8\linewidth]{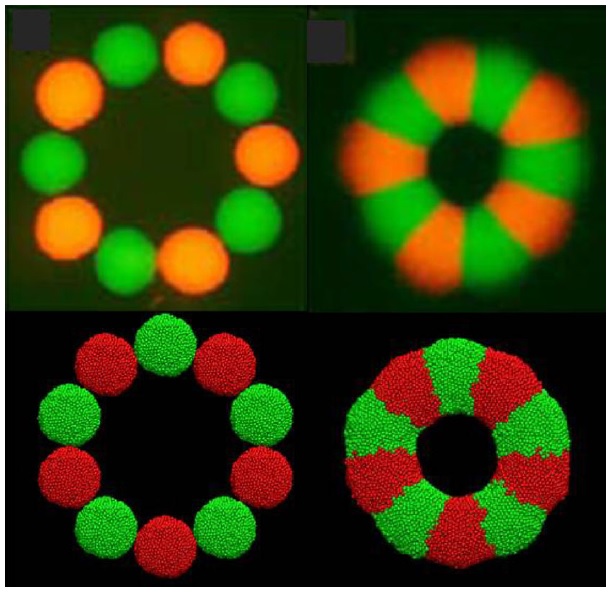}
\caption{ Cellular torus build with CHO spheroids. Top raw shows the experiment (\cite{Jakab2008}) results both initial state (top left) and final results (top right). And lower raw shows the simulation results using CPD method \cite{Kosztin2012}.  }
\label{fig:CPD_Experiment}
\end{figure}

   Where $f_{i,n} (t)$ stands for the fluctuations in the dynamics of the cellular cytoskeleton. This force is modeled as a Gaussian white noise with zero mean and variance $<f_i (t)f_j (0)>=2D\mu ^2 \delta(t) \delta_{ij}$, where D is the sort-time self-diffusion coefficient of the CPs. The CPD parameters D and $\mu$ are related to the previously introduced biological fluctuation energy $E_T$ by the Einstein relation $D\mu= E_T$ \cite{Flenner2008,Flenner2012}. Implementing CPD into molecular dynamics (MD) package, Kosztin et al. \cite{Kosztin2012} were able to verify the model with experimental results (Figure \ref{fig:CPD_Experiment}). \\
   
  \subsection{Printing techniques} 
    
  Having the corrected version of the three-dimensional geometry of the tissue, it is time to process the generated data and used them to print out the tissue/organ. Considering both wide categories of the bio-printers (laser-assisted and laser free methods), here we will briefly review available bio-printers in both categories.
  
  \subsubsection{Laser assisted bio-printers (LaBP)}
  
Laser assisted printers were first introduced by Odde and Renn \cite{Odde2000a} as the first bio-printer under title of “Laser-Guided Direct Writing” (LG DW) \cite{Bakhshinejad2015}. 

Despite its high accuracy, it was modified soon after the announcement because of the low cell viability \cite{Tasoglu2013}; new method called “Laser-induced forward Transfer (LIFT)” . The second method (LIFT) was still a direct-writing technique with a small change in the way of writing. In this method laser used to vaporized small patch of bio-ink in order to generate droplets  \cite{Zergioti2005,Zergioti2005a,wu2003}. 

Due to direct contact of the cell with the laser and the amount of transferred energy to the cell, the viability of the cell was low. So, with a slight modification of the method and adding a thin layer of absorption, this method was able to improve the viability to more than 95\% while kept the accuracy and resolution of LIFT \cite{Ringeisen,Barron2004a,Chen2006}

\begin{figure}[t]
\centering\includegraphics[width=0.8\linewidth]{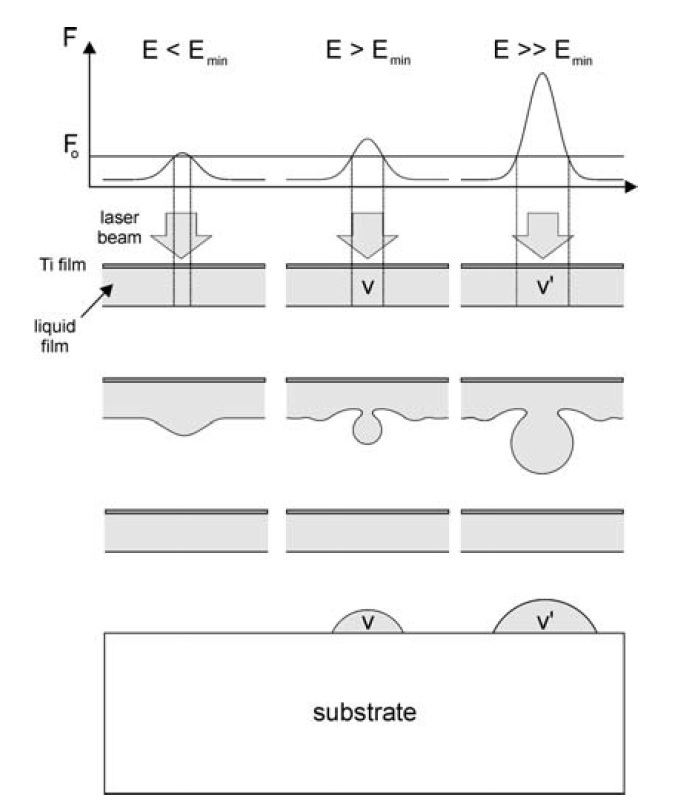}
\caption{ Schematic representation of the ejection process using LaBP in response to different energy signals. Image adopted from \cite{Serra2006}}
\label{fig:LaserEnergy}
\end{figure}

All these methods are initially developed based on the principles of laser-induced forward transfer which initially developed to transfer metals \cite{Barron2004,Bohandy1986,Yan2013,Ringeisen}. A typical LaBP device, consists of a pulsed laser beam, a focusing system, a ‘ribbon’ which is usually a glass covered by laser energy observing layer and a layer of biological material and finally a receiving substrate. One of the factors that effects on the resolution of the LaBP devices is the laser fluence. Depending on the amount of energy and time period of the fluence, size of the droplet can be vary (Figure \ref{fig:LaserEnergy}). There are many other factors that affect the resolution (e.g., surface tension, viscosity and etc.). A LaBP device can reach to resolution of a single cell per droplet with density of up to $10^8$ cells/ml \cite{Murphy2014,Guillotin2010}. The biggest disadvantage of this method is the usage of the ribbon as the primary part which make it hard to use in scaffold free tissue printing process. There are two main problems associating with using ribbon; first, for each cell type there should be an individual ribbon coated with that cell type, on the other hand, since there is no pre-knowledge about the location of the cells on the ribbon we might end up with having cell free droplets. The second problem can be solved by using one of the image processing techniques (i.e., cell recognition) in order to locate the cells. \\

Considering all advantages and disadvantages of LaBP devices, they are the main tools for fields like scaffold development \cite{Nair2008,Hutmacher,Ahn2014}, lab on the chip \cite{Gunther2010,Hribar2014} etc. but not for the scaffold-free tissue printing because of the nature of these processes which requires printing of different cell types at nearly same time. 

\subsubsection{Laser free bio-printers (LfBP)}

Due to the complexity of the LaBP devices, and their big disadvantage of not being able to print more than one cell type at a time, different groups of bio-printers have been developed to overcome this difficulties. These devices, Laser free Bio-Printers (LfBP), can be categorized into two wide categories: Inkjet bio-printer and robotic dispensing bio-printers \cite{Murphy2014}.  The printing process is the same for all methods and consists of a printer head, an ink cartridge/reservoir and a computer-controlled receiving substrate \cite{Ringeisen2006}.   All these methods have overcome the single cell type printing problem by using parallel (multiple) printer heads which can be seeded with different cell types. Printer heads and cartridges were modified in order to be capable of using bio-ink for printing \cite{Pardo2003}. The bio-inks usually consists of aqueous media, thermo-reversible polymers, or polymer/hydrogel precursors combined which seeded with living cells.  Like any other method, these methods have their own disadvantages. LfBP’s devices biggest disadvantage is the clogging problem. Another problem that comes from this methods is that, to best of our knowledge, there is no published paper about controlling the number of cells per droplets for LfBP devices. 

There are three types of ink-jet bio-printers that have been developed: Thermal, Piezoelectric and valve-based methods. Wilson and Boland \cite{Wilson2003} reported the first viable cell printed by modified thermal print head. Mironov et al. \cite{Mironov2003} and Boland et al. \cite{Boland2003} reported future developments of the same device.  In a thermal printer head, a micro-heater element vaporizes a small pocket of fluid, the formation and collapse of the vapor bubble generates an acoustic pressure pulse. The thermal pulse causes transient membrane pores that close in a course of several hours \cite{Cui2010,Xu2009}.  In Piezoelectric approach, no thermal pulse applies to the bio-ink, instead mechanical deformation of the chamber causes discontinuity in the bi-ink and generates droplets \cite{Malda2013}. The valve-base method is uses multiple solenoid valves (depends on the number of different cells) which is controlled by a microcontroller system to generate droplets \cite{Faulkner-Jones2013}. All mentioned ink-jet as well as LaBP bio-printers are also referred as drop-on-demand methods (DoD), since they generate droplets at required locations \cite{Derby2008}. The last group of bio-printers is the robotic dispensing printer. In this approach, the bio-ink is extruded as continuous beads of bio-ink instead of generating droplets using robotically controlled extrusion system. The most common methods for this category are pneumatic or mechanical (piston or screw) systems \cite{Murphy2014,Khalil2007,Chang2011,Jakab2006,Visser2013}. Pneumatically driven systems have the advantage of having simpler drive-mechanism components where the mechanical driven system are given more direct control over the material flow (Figure \ref{fig:Bio-printing_steps}). 

\section{Conclusion}

Scaffold-free bio-printing is one of the most promising and challenging methods in tissue engineering. Like any other production process, this method also needs a process plan in order to successfully print-out the tissue/organ. Since bio-printing involves post-printing processes which cause changes in the shape and properties of the product, in scaffold-free methods, a predicting model is essential. Different mathematical models have been developed in order to predict the steady state of the bio-printing process. As a results, CPD and CPM were reported as the most accurate models. In a closer look, CPD is capable of modeling larger number of cells where as CPM is able to model less number but in more details. However, due to the lack of experimental technologies where there is no possible way to print out solid tissues with a thickness of more than a few hundred micrometers, in order to evaluate the simulated results, there is not any reported result for simulating the whole process. This predicting model can be done by using of the CPD model which requires a relatively expensive process. There are developments that have to be done in the future for every step of the process, both for biologists and physicists in the field.  Issues that has to be addressed by physicists are bio-printer technology and mathematical models. Bio-printing technologies that have been mentioned in this review still have issues for commercialization in terms of speed. Also these methods have not reached the required resolution \cite{Murphy2014}.  However, as we mentioned earlier, as the experimental results are growing, larger scale models are required with a need for finer details which can model the maximum possible number of reactions and post-printing processes for the whole organ.



%
% ...

\bibliographystyle{asmems4}

\bibliography{bib}


\end{document}